\documentclass[reprint,amsmath,amssymb,aps]{revtex4-2}

\usepackage{graphicx}
\usepackage{dcolumn}
\usepackage{bm}
\usepackage{float}
\usepackage{amsmath}

\usepackage{xcolor}

\begin{document}

\title{Hierarchical structure of the energy landscape in the Voronoi model of dense tissue}

\author{Diogo E. P. Pinto$^{1,2}$, Daniel M. Sussman$^{3}$, Margarida M. Telo da Gama$^{1,2}$ and Nuno A. M. Ara\'{u}jo$^{1,2}$}

\affiliation{$^{1}$ Centro de Física Teórica e Computacional, Faculdade de Ciências, Universidade de Lisboa, 1749-016 Lisboa, Portugal. \\ $^{2}$ Departamento de Física, Faculdade de Ciências, Universidade de Lisboa, 1749-016 Lisboa, Portugal.  \\ $^{3}$ Department of Physics, Emory University, Atlanta, GA, USA}

\begin{abstract}
The Voronoi model is a popular tool for studying confluent living tissues. It exhibits an anomalous glassy behavior even at very low temperatures or weak active self-propulsion, and at zero temperature the model exhibits a disordered solid structure with no evidence of a rigidity transition. Here we investigate the properties of the energy landscape in this limit. We find two disordered solid phases that have similar structural features but that differ in the ultrametricity of their energy landscapes; the crossover between these two states shares phenomenological properties with a Gardner transition. We further highlight how the metric used to calculate distances between configurations influences the ability to detect hierarchical arrangements of basins in the energy landscape.
\end{abstract}

\maketitle

\begin{figure}[t]
	\includegraphics{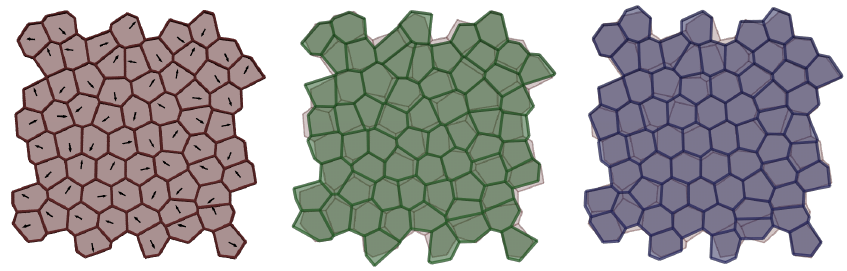}
	\caption{\label{model} Representation of the perturbation protocol. On the left is represented the original minimized configuration (red) with the perturbation vectors in the center of each cell. In the middle is the original minimized configuration (red) and the perturbed one (green). On the right is the original minimized configuration (red) and the one minimized after the perturbation (blue).}
\end{figure}

Understanding the collective behavior of cells in biological tissues has become one of the major interdisciplinary challenges of recent years, with applications ranging from wound healing to cancer treatment~\cite{Ghosh2007, Gov2009, Park2015, Tambe2011, PerezGonzalez2018, Sunyer2016}. Both experimental and theoretical efforts have been crucial in understanding the properties of these tissues and the mechanisms by which tissue properties are regulated, for example in the way that tissues can transition from rigid to flexible as the properties of individual cells are regulated~\cite{Angelini2011, Garcia2015, Park2015, Schotz2013, mongera2018fluid,Grosser2021,devany2021cell}. Rigidity transitions are also seen in particulate systems, such as granular materials and colloidal suspensions, in which changes in particle density and temperature can lead to disordered materials in a kinetically arrested jammed or glassy state~\cite{Binder1986, Berthier2011, Janssen2019}. In living tissues, the nature and properties of the rigid states are still under debate~\cite{Angelini2011, Garcia2015, Park2015, Schotz2013}, owing both to the explicitly non-equilibrium nature of cellular motion and the many-body interactions found in confluent tissue~\cite{Szavo2006}. These differences raise several challenges to the generalization of ideas and methods developed in the context well-studied particulate matter~\cite{Merkel2018, Merkel2019, Angelini2011, Tambe2011, PerezGonzalez2018}.

Several models have been proposed to understand the collective behavior of cellular systems, from single particle descriptions to density field models~\cite{Bi2014, Bi2015, Farhadifar2007, Fletcher2014, Camley2017, Kabla2012}. The Voronoi model represents a confluent tissue as a space filling polygonal tiling, where each positional degree of freedom corresponds to a cell whose shape is obtained by an instantaneous Voronoi tessellation~\cite{Bi2016,Kaliman2016, Sussman2018a}. The dynamics is controlled by an energy functional that is quadratic in the area and perimeter of each cell (described in more detail below), and the mechanical properties of the tissue can be either solid-like or fluid-like depending on the temperature and a shape parameter, $p_0$, which quantifies the target shape of the individual cells~\cite{Bi2016}.

At zero temperature (or in the absence of cellular activity) it has been argued that the Voronoi model possesses a finite shear modulus over its entire range of model parameters~\cite{Sussman2018}. This is in sharp contrast with particulate systems, in which a zero-temperature rigidity transition can be observed by changing the system density~\cite{Parisi2010, Goodrich2012, Charbonneau2017, Baule2018}. The particulate jamming transition is typically interpreted in the context of constraint counting, in which the transition occurs when the number of independent particle-particle contacts equals the number of degrees of freedom~\cite{Lubensky2015,franz2019critical}. As described below, the 2D Voronoi model is always at this point of marginal stability~\cite{Sussman2018}, and thus an analysis based only on the balance between constraints and degrees of freedom is insufficient. It has been proposed instead that energetic rigidity, in which not only simple constraints but also residual stresses play a controlling role~\cite{Merkel2018, Merkel2019,Yan2019, Damavandi2021}, is a better framework for understanding Voronoi model rigidity in the athermal limit~\cite{Damavandi2021}.

\begin{figure*}[t]
	\includegraphics{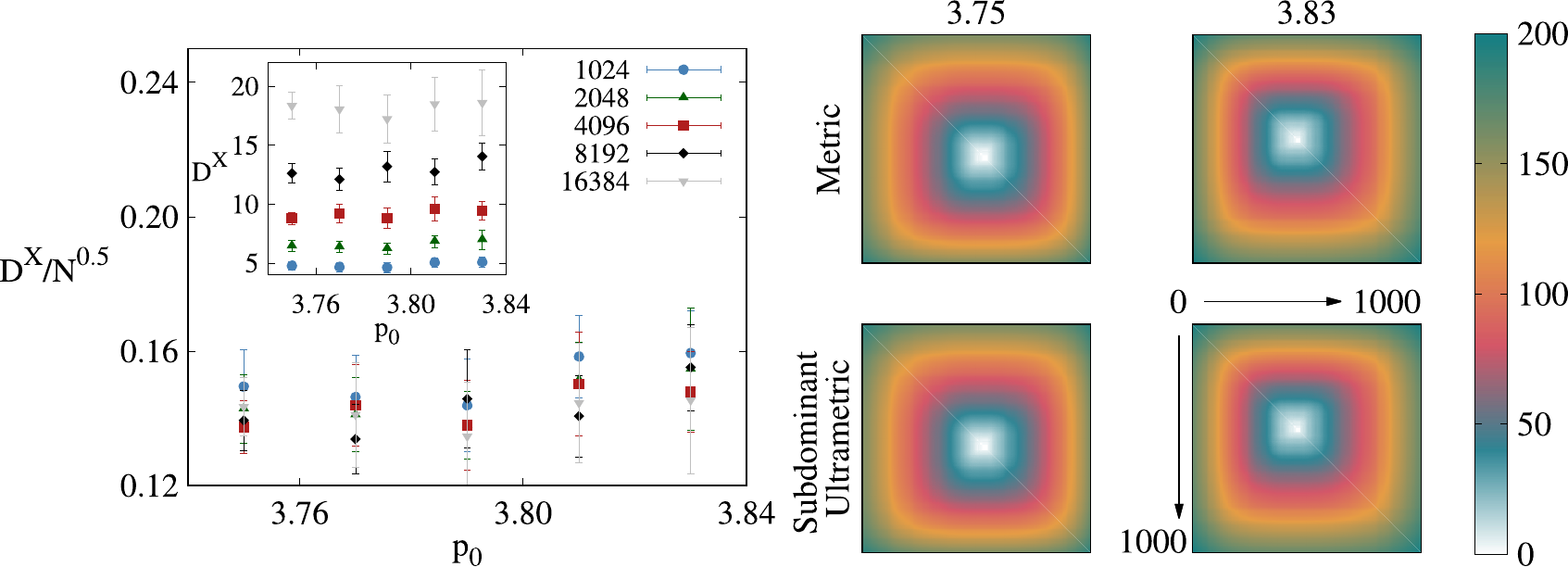}
	\caption{\label{PosMetric} (Left) The normalized generalized distance to ultrametricity as measured using the contact vector metric, $D^X/\sqrt{N}$, as a function of  $p_0$, for $N=1024, 2048, 4096, 8192, 16384$. The inset shows the same results without the scaling. (Right) A schematic representation of the distances between minima according to the contact metric, $d^X(a,b)$, and the subdominant ultrametric constructed from it using a minimum spanning tree~\cite{Kruskal1956, Rammal1985}. Matrices corresponding to $N=4096$ and $p_0 = 3.75,\ 3.83$ are shown, where the different distances are grouped using a single-linkage clustering algorithm which clusters the minima sequentially by distance. All results are averages of 10 initial configurations subject to 100 perturbations and minimizations each.}
\end{figure*}

Here we explore this unusual athermal regime of the Voronoi model and show that, even in the absence of a zero-temperature rigidity transition, there is nevertheless a profound change in the statistics of the energy landscape in different regions of the model parameter space. We find evidence for a transition to an ultrametric, hierarchical arrangement of basins in the energy landscape, suggesting two different phases of a disordered solid~\cite{Charbonneau2015, Liao2019, Artiaco2020, Dennis2020}. The ultrametric state is characterized by energy minima forming a tree-like structure in phase space where minima within a given sub-basin are much closer to one another than they are to minima in any other sub-basin~\cite{Parisi2010, Charbonneau2017}. This is consistent with a Gardner phenomenology~\cite{Parisi2010, Charbonneau2017}; the phenomenological properties of this phase have been studied in multiple experimental and computational systems~\cite{Charbonneau2015,seguin2016experimental,scalliet2017absence,Liao2019, Artiaco2020, Dennis2020}.

We model the confluent tissue as a monolayer of $N$ cells~\cite{Bi2016,Bi2014,Bi2015} in a square domain of side-length $L$ with periodic boundary conditions. Each cell $i$ is represented by its center $\textbf{r}_i$ with a shape given by an instantaneous Voronoi tessellation of the space. We choose the unit of length to be given by the square root of the average area of all the cells. We can then write a dimensionless version of the contribution of each cell to the energy functional as~\cite{Farhadifar2007,Fletcher2014,Merkel2018,Teomy2018a}
\begin{equation}\label{eq::en_func_reduced}
	e_i=k_A(a_i-1)^2+(p_i-p_{0})^2.
\end{equation}
Here $a_i$ and $p_i$ are the dimensionless area and perimeter of cell $i$, $p_0$ is the target perimeter, and $k_A$ represents the ratio between the relative stiffness of the area and perimeter elasticity of the cell. Biologically, the first term models cellular incompressibility and the resistance of the cellular monolayer to height fluctuations; the second term models the competition between active contractility of the actomyosin subcellular cortex and the effective cell membrane tension due to cell-cell adhesion and cortical tension.

For each set of parameters, we start with $N$ cells distributed at random positions. The configuration is then minimized using the FIRE algorithm~\cite{Bitzek2006, Sussman2017}, which we halt when the maximum net force on each cell is less than $10^{-12}$. Due to numerical constraints, we consider $p_0\leqslant3.85$, since it has been shown for athermal systems that, above this value, configurations with multi-fold vertices are obtained which lead to numerical instabilities in the minimization protocol~\cite{Sussman2018}. Further details of the simulations can be found in the \textit{Supplemental Material}.

To probe the structure of the energy landscape, we start from an initial configuration that corresponds to a local minimum and perturb it to find new stable configurations. In our primary perturbation protocol, we displace the position of each cell according to the vector $\overrightarrow{P}_{\varepsilon}=[X_0, X_1,\ldots, X_{2N-1}]$, where $N$ is the number of cells, $X_{2i}=\varepsilon\cos(\theta_i)$ is the perturbation to cell $i$ along the $x$-axis, $X_{2i+1}=\varepsilon\sin(\theta_i)$ is that along the $y$-axis, and $\theta_i$ is a random angle uniformly distributed between $0$ and $2\pi$. We considered a random length $\varepsilon$ drawn from a uniform distribution between $0$ and $\varepsilon_{max}$, to guarantee the possibility of visiting minima in the same and different top-level basins. The norm of the perturbation vector is $|\overrightarrow{P}_{\varepsilon}|=P_{\varepsilon}=\varepsilon\sqrt{N}$. After the perturbation, we subtract the global translation of the tissue, and then let the tissue relax to a new minimum. An example of this process is shown in Fig.~\ref{model}; details of alternate ``perturb-and-minimize'' schemes can be found in the \textit{Supplemental Material}, where we show that our results are not qualitatively sensitive to these details.

\begin{figure*}[t!]
	\includegraphics{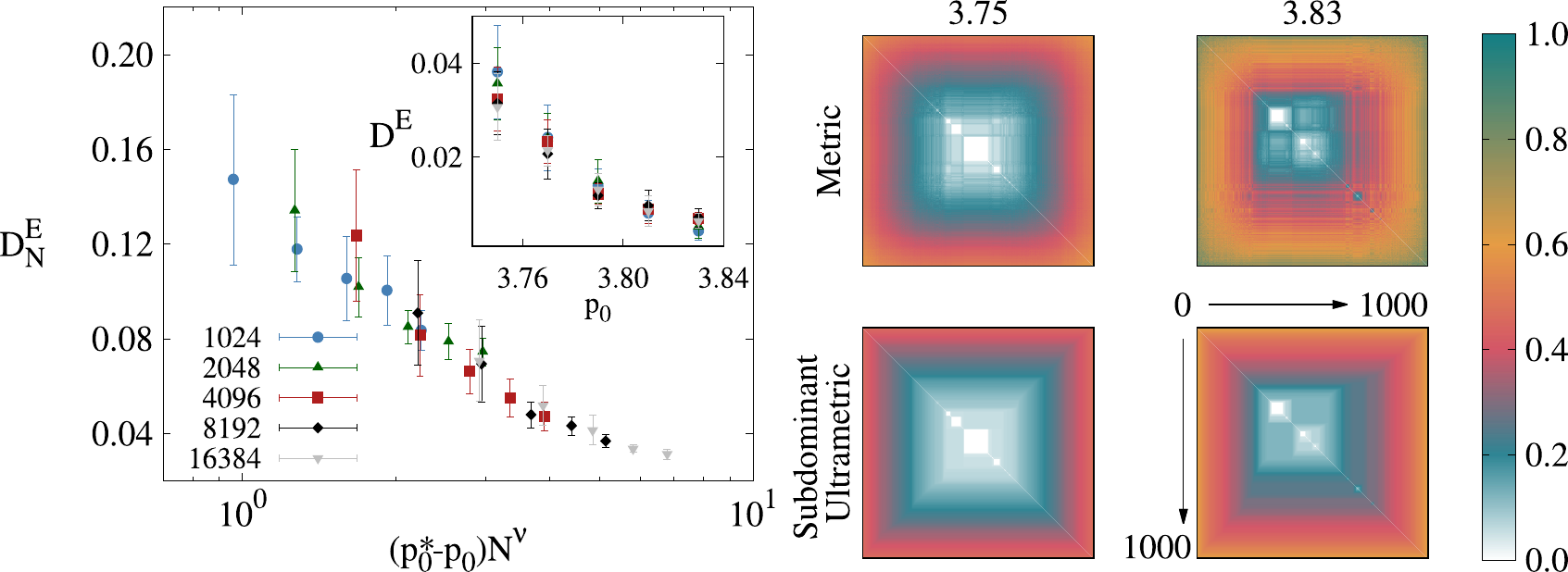}
	\caption{\label{EnergyMetric} (Left) The normalized generalized distance to ultrametricity as measured using the normalized energy metric, $D^E_N$,  as a function of $(p_0^*-p_0)N^{\nu}$, for $N=1024, 2048, 4096, 8192, 16384$, $p_0^*=3.89\pm 0.01$ and $\nu=0.4\pm 0.01$. The inset shows the generalized distance to ultrametricity, $D^E$, calculated using the energy metric, $d^E(a,b)$, as a function of $p_0$, for the same $N$. (Right) Schematic matrix representation of the distances between minima according to the normalized energy metric, $d_N^E(a,b)$, and its subdominant ultrametric, as in Fig.~\ref{PosMetric}. Matrices are shown for $p_0=3.75,\ 3.83$ and $N=4096$. All results are averages of 10 initial configurations subject to 100 perturbations and minimizations each.
	}
\end{figure*}

We will be exploring different metrics to characterize distances between minima in the energy landscape. To begin quantifying these distances we consider the contact metric (denoted by the superscripted $X$) discussed in Refs.~\cite{Dennis2020, Liao2019, Artiaco2020},
\begin{equation}\label{metric}
	d^X(a,b)=\sqrt{\sum_{ij}(\vec{C}^{a}_{ij}-\vec{C}^{b}_{ij})^2},
\end{equation}
where $d^X(a,b)$ is the distance between configuration $a$ and $b$, and $\vec{C}^{a}_{ij}$ is the 2D contact vector between two cells, where each component $C^{a}_{ij,x}=x_i^a-x_j^a$ is the distance along the respective axis between cells $i$ and $j$ if those cells share and edge, and $\vec{C}^{a}_{ij}=\vec{0}$ otherwise. In the \textit{Supplemental Material}, we show how the normalized distance to the original minimum, $d^X_N(a,b)=d^X(a,b)/\sqrt{|a| |b|}$, where the norm of a configuration corresponds to $|a|=d^X(a,0)$, scales with the number of cells and the norm of the perturbation vector. We observe that $d^X(a,b)\sim\sqrt{N}$ and choose $\varepsilon_{max}=0.5\sqrt{N}$ for all $p_{0}$, which is large enough so that the perturbed configuration does not always relax to the initial minimum but small enough that nearby minima are accessible.

Figure \ref{PosMetric} depicts matrices where the color corresponds to the distance between minima, for all pairs of minima found for $p_0=3.75$ and $3.83$. These matrices were constructed for $10^3$ minima obtained for a tissue of $4096$ cells. Each element of the matrix corresponds to the distance between two minima, $a$ and $b$, given by Eq.~\eqref{metric}. The distances are all sorted using the single-linkage clustering algorithm on the metric of the fully minimized systems, which groups them sequentially based on their relative distance~\cite{Murtagh1983}. The colors represent different distances. In white are the minima that are closest to each other. We find groups of minima that are all at this minimum distance, forming a white region that corresponds to sub-basins. The matrices do not show any substantial visual change with $p_0$.

A Gardner phase is characterized by an ultrametric phase space consisting of a tree-like structure, where minima within a given sub-basin are all much closer to one another than to minima in any other sub-basin. This property is codified by an ultrametric inequality,
\begin{equation}
\label{ultraIneq}
d^X(a,c)\leq \text{max}[d^X(a,b) , d^X(b,c)] ,
\end{equation}
where $a, b$ and $c$ are three different configurations in phase space and $d^X(a,b)$ is the distance between configurations. To verify if the properties of the tissue are consistent with a Gardner phase, we compute how close the metric is to being ultrametric. To do so, we first find the subdominant ultrametric, $d^<(a,b)$, i.e., the ultrametric that is closest to $d^X(a,b)$ itself. The subdominant ultrametric can be found by first computing the distances in the minimum spanning tree of the space of minima. Then, for each pair of minima $a$ and $b$, we compute the path between them in the minimum spanning tree and define $d^<(a,b)$ as the largest distance between two neighboring minima along the path~\citep{Rammal1985}. The corresponding matrices are shown in Fig.~\ref{PosMetric}.

Having found $d^<(a,b)$, we finally calculate the generalized distance between the metric and the subdominant ultrametric using
\begin{equation}\label{genDist}
	D^X=\sqrt{\langle (d^X(a,b)-d^<(a,b))^2 \rangle},
\end{equation}
where $\langle \cdot \rangle$ denotes the average over all configuration pairs $a$ and $b$. If $D^X=0$ then the energy landscape is ultrametric, while $D^X>0$ quantifies how far it is from ultrametricity. In the inset of Fig.~\ref{PosMetric}, we show that $D^X$ increases slightly with $p_0$, but more importantly it depends strongly on $N$. In the main plot we re-scale $D^X/\sqrt{N}$ and obtain a reasonable collapse of the data. Since the typical distance between minima and the distance to ultrametricity both scale with $\sqrt{N}$, the space with this contact metric is not ultrametric in the thermodynamic limit~\cite{Dennis2020}.

Using distances based on the contact vectors suggests that the landscape of the Voronoi model is not ultrametric, but does the choice of metric itself influence this result? We note that in the Voronoi model, the contact network of the tissue does not change significantly even as the tissue rigidity changes substantially~\cite{Damavandi2021, Merkel2019}. We further note that the energy functional in Eq.~\eqref{eq::en_func_reduced} is a simple collection of harmonic springs, in a coordinate basis of shape space rather than in the positional basis of the degrees of freedom generating the shapes. This suggests a different metric might be more appropriate, and in this context we propose one based on the contribution of each cell $i$ to the total energy of the tissue, $E_i$. We take the same form for the metric as Eq.~\eqref{metric}, but where $\vec{C}^a_{ij}\rightarrow C^a_{ij}=E^a_i-E^a_j$, if $i$ and $j$ are neighbors and zero otherwise. We call this metric the ``energy metric'', $d^E(a,b)$. We adopt the same perturb-and-minimize protocol as before, using  $\varepsilon_{max}=0.1\sqrt{N}$ for all $p_{0}$. As such, we keep biasing the perturbations to nearby minima in the new metric, otherwise more simulations would be needed to probe the same volume of configuration space. In the \textit{Supplemental Material} we show that the results using the contact metric remain qualitatively the same using the new  $\varepsilon_{max}$. Just like the contact metric, the energy metric scales with system size, $d^E(a,b)\sim\sqrt{N}$, since it depends on the total number of cell-cell contacts.  We observe in the inset of Fig. \ref{EnergyMetric} that, for the energy metric, the distance to ultrametricity ($D^E$) does not scale with $N$ and decreases with $p_0$ (inset of Fig.~\ref{EnergyMetric}). Since the distance between minima scales as $d^E(a,b)\sim\sqrt{N}$, while the generalized distance does not depend on the system size, this suggests that the system does, in fact, become ultrametric in the thermodynamic limit: $D^E/d^E(a,b)\sim 1/\sqrt{N}$.

In the case of the contact metric the calculated values are already normalized since we increase the box size with $N$, while the typical cell size is fixed. In the case of the energy metric this is no longer the case. Thus, to properly compare the system properties at different $p_0$ (since $\langle E_i\rangle$ varies with $p_0$), we also consider a normalized version of the energy metric: $d^E_N(a,b)=d^E(a,b)/\sqrt{|a| |b|}$, for which the typical distance between configurations does not depend on either $p_0$ or $N$. Figure \ref{EnergyMetric} shows the schematic representation of the normalized energy metric, $d_N^E(a,b)$, and its subdominant ultrametric. From this representation, we can already observe changes in the structure of the energy landscape. As $p_0$ increases, the color gradient is less smooth and the boxes corresponding to the different sub-basins become sharper. Furthermore, it is also observed that more minima fall into the same sub-basin. These properties suggest that, when the value of $p_0$ decreases, the structure of the energy landscape becomes more hierarchical~\citep{Dennis2020}.

We also compute the generalized distance to ultrametricity when using $d_N^E$. Again we see that ultrametricity is approached with increasing system size. Furthermore, with the scaling shown in the main panel of  Fig.~\ref{EnergyMetric} we can collapse the different curves. The values of $p_0^*$ and $\nu$ were chosen such that we obtain the best collapse. Recent work in particulate systems has interpreted a similar scaling as a distance to jamming~\cite{Dennis2020, Goodrich2012}. This suggests not only that the Voronoi model has an ultrametric structure in the thermodynamic limit, but that there may be a transition between two different solid phases. Previously, it was shown that the athermal Voronoi model did not have a rigidity transition~\cite{Sussman2018}. Nevertheless, we show that there is a clear difference between the energy landscape for low and high $p_0$: at low $p_0$ the glass state is characterized by large residual stresses and an ultrametric energy landscape, and at high $p_0$ the energy landscape is not ultrametric. We find that the change in the structure of the energy landscape occurs for preferred shape parameters in the range $p_0=3.75-3.83$, which is close to where the zero-temperature shear modulus changes markedly~\cite{Sussman2018} and where the dynamics at finite temperature changes in character~\cite{Bi2016, Sussman2018a,Li2021}.

\begin{figure}[t!]
	\includegraphics{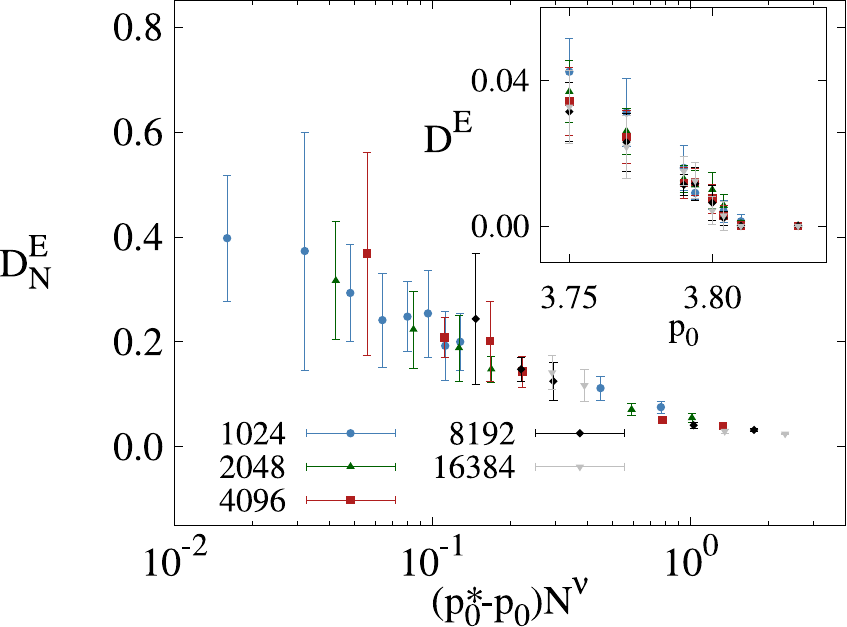}
	\caption{\label{EnergyKAMetric} A plot as in Fig.~\ref{EnergyMetric} but for $k_A=0$, highlighting the similar scaling in the two cases. Here, we use $p_0^*=3.798\pm0.001$ and $\nu=0.4\pm 0.01$. All results are averages of 10 initial configurations subject to 100 perturbations and minimizations each.
	}
\end{figure}

As $p_0$ increases, fewer sub-basins are found inside each basin (as represented by the different unconnected clusters in Fig.~\ref{EnergyMetric}), suggesting that the energy landscape flattens out. Another way of exploring this flattening of the energy landscape is by studying the behavior of the model as the relative area modulus $k_A$ is varied. In the limit $k_A=0$ the Voronoi model is no longer marginally constrained, and it acquires a residual-stress-based rigidity transition as a function of $p_0$ at $T=0$~\cite{Sussman2018}. As shown in Fig.~\ref{EnergyKAMetric}, for $p_0<3.79$ we find that the tissue is both rigid and the energy landscape is ultrametric. For slightly larger $p_0$ the energy landscape deviates from ultrametricity and the variance of $D_E^N$ increases significantly. In this regime, the energy landscape consists of a mixture of hierarchical basins and several nearly flat basins and so, in many cases, small perturbations will not drive the system to a different minimum. Due to finite size effects it is difficult to assess if a new solid phase exists. In the \textit{Supplemental Material} we use a simple technique to estimate the transition from the solid to the fluid phase. Using the fraction of configurations with zero energy we estimate a transition point around $p_0^*=3.8022\pm0.0001$, while in Fig.~\ref{EnergyKAMetric} $p_0^*=3.798\pm 0.001$ seems to be more appropriate value to collapse the data. We used $\nu=0.4\pm 0.01$, which was the best value for the collapse. More simulations would be needed to conclude whether a new solid phase at $k_A=0$ exists before the fluid phase, or if the observations are finite size effects and both transitions actually coincide. For $p_0>3.8$ the energy landscape is flat, characteristic of a fluid-like tissue. For any $k_A>0$, different energy minima are found for all $p_0$, consistent with previous work showing a finite shear modulus for the whole range of model parameters investigated~\cite{Sussman2018}. Although we have not explored the thermal case, recent studies with the thermal 2D Voronoi model also suggest a change in the energy landscape close to $p_0\approx3.81$, where there is an emergence of a fractal-like energy landscape and cells become virtually free to diffuse in specific phase space directions up to a small distance~\cite{Li2021}.

In summary, we have found indications of a hierarchical structure of the energy landscape in a model of dense biological tissue whose zero-temperature rigidity is quite different from that of constraint-based particulate systems. Strikingly, we find that the choice of metric to characterize distances between minima is crucial: defining distance based on changes in neighboring contact vectors vs contributions to the energy give \emph{qualitatively} different interpretations of the structure of the energy landscape. In particulate systems the contact vectors enter explicitly in the relevant energy functional -- i.e., the energy of a soft harmonic repulsion or a Lennard-Jones interaction is a simple function of the contact vector between interacting particles. In Voronoi models the energy cannot be decomposed into independent pairwise contributions, which we speculate is the reason that choosing a distance metric based on the total energy associated with each degree of freedom is required to uncover the hierarchical structure of the landscape. We further speculate that this may point more generally to the importance of the choice of metric for systems in which many-body interactions dominate over pairwise ones. An avenue for future research could be relating these tissue-like systems to particulate ones, such as soft or hard spheres~\cite{Liao2019, Dennis2020}. This could be done by establishing the relation between the effects of $p_0$ in the Voronoi model and pressure in particulate systems. Since both exhibit an ultrametric landscape, this could allow a generalization of glassy physics outside of particulate systems and glass-forming materials~\cite{Berthier2011, Parisi2010, Janssen2019}.


\section{Acknowledgments}

The authors acknowledge financial support from the Portuguese Foundation for
Science and Technology (FCT) under Contracts no. PTDC/FIS-MAC/28146/2017
(LISBOA-01-0145-FEDER-028146), UIDB/00618/2020, UIDP/00618/2020 and
SFRH/BD/131158/2017.

\bibliography{Glass}

\end{document}